\documentclass[aps,floatfix,showpacs,superscriptaddress,preprint]{revtex4}

\usepackage{latexsym,graphicx,epsfig,psfrag,fancybox,xspace}
\usepackage{amssymb,amsmath,amsxtra,amsfonts}
\usepackage{bm}

\usepackage{longtable}
\usepackage{multirow}
\usepackage{booktabs}
\usepackage{array}
\usepackage{wrapfig}
\usepackage{color}
\usepackage{titlesec}

\titleformat{\section}{\large\bfseries}{\thesection}{1em}{}

\newcommand{\bea}{\begin{eqnarray}}
\newcommand{\ena}{\end{eqnarray}}
\newcommand{\be}{\begin{equation}}
\newcommand{\en}{\end{equation}}
\newcommand{\nn}{\nonumber\\}
\newcommand{\ed}{\end{document}} 

\newcommand{\ord}{\mathcal{O}}

\begin{document}

\title{Angular observables and branching ratio for $B_s\to \phi \ell^+ \ell^-$ decay} 

\author{Aidos Issadykov}
\email{issadykov.a@gmail.com}

\affiliation{The Institute of Nuclear Physics, \\
Agency of the Republic of Kazakhstan for Atomic Energy, 1 Ibragimova, 050032, Almaty,  {\it Kazakhstan}}

\affiliation{
Joint Institute for Nuclear Research, 6
Joliot-Curie, 141980, Dubna, Moscow region, \textit{Russia}}

\begin{abstract}
 An analysis of the $B_s\to \phi \ell^+ \ell^-$ rare decay within the covariant confined quark model is performed in this article.  The $B_s\to \phi$ transition form factors are calculated and then used to compute the branching fractions and angular observables in various $q^2$ bins, including the forward-backward asymmetry $A_{FB}$, the longitudinal polarization $F_L$, and the optimized observables $A_i$ and $S_i$.  Angular observables and branchings were calculated for the bins that were presented by the LHCb collaboration.
 Generally CCQM predictions are consistent with data from LHCb collaboration within the error bars. However, the largest discrepancy is observed in the forward-backward asymmetry of $A_{FB}$ for the bin [6,8]~GeV$^2$ due to the reverse signs. Discrepancies are also observed in the observed values of $A_{FB}$ and $A_5$ for the bins [11,12.5] and [15,18.9]~GeV$^2$.

\end{abstract}


\maketitle
\section{Introduction }

Among the $b\to s\ell^+\ell^-$ transitions, the rare decays $B\to K^\ast(\to K\pi)\mu^+\mu^-$ and $B_s\to \phi(\to K^+K^-)\mu^+\mu^-$ are the most prominent and extensively studied. The first angular analysis of $ B^0_s\to \phi \mu^+\mu^-$ decay by LHCb \cite{Aaij:2015esa} found results largely consistent with the Standard Model (SM), though the measured decay width in one bin was more than 3~$\sigma$ below SM predictions. These measurements were subsequently refined in 2021, with a new analysis reporting updated branching ratios, branching fractions, and angular observables in bins \cite{LHCb:2021zwz,LHCb:2021xxq}.

\begin{align}
  \frac{ {\cal B}(B_{s}^{0} \to \phi \mu^{+} \mu^{-}) }{ {\cal B}(B_{s}^{0} \to J/\psi \phi) } & = (8.00 \pm 0.21 \pm 0.16 \pm 0.03) \times 10^{-4},
  \nn
     {\mathcal B} (B_{s}^{0} \to \phi \mu^{+} \mu^{-}) & = (8.14 \pm 0.21 \pm 0.16 \pm 0.03 \pm 0.39)\times 10^{-7}.
\end{align}

The experimental investigation of the $B_s \to \phi \mu^+\mu^-$ decay began with its observation by the CDF collaboration \cite{CDF:2011grz} and has since been extensively refined by LHCb \cite{LHCb:2013tgx, Aaij:2015esa,LHCb:2021zwz,LHCb:2021xxq}. These studies have progressed beyond simple branching ratio measurements to a detailed analysis of the decay's full kinematic phenomenology.

A wide range of theoretical studies have been performed on $B_s \to \phi \mu^+\mu^-$ exclusive decays. Within the Standard Model, these include the covariant and light-front quark models ~\cite{Deandrea:2001qs,Dubnicka:2016nyy,Geng:2003su}, QCD factorization~\cite{Bobeth:2008ij}, and light-cone sum rules \cite{Ball:2004rg,Altmannshofer:2014rta,Wu:2006rd,Straub:2015ica,Descotes-Genon:2015uva}. New physics scenarios, such as universal extra dimensions~\cite{Li:2011yn} and supersymmetry~\cite{Xu:2013lms}, have also been investigated.
The lattice QCD form factors from \cite{Horgan:2013hoa} have been employed in several studies. They were used in \cite{Horgan:2013pva, Horgan:2015vla} to calculate the decay widths for $B\to K^\ast\mu^+\mu^-$ and $B_s\to \phi \mu^+\mu^-$. A different approach, the PQCD factorization method, was also applied using this lattice input in \cite{Jin:2020qfp}.

\section{\boldmath{$B_s\to\phi$ transition form factors}}
\begin{figure*}[htbp]
\centering
\begin{tabular}{lr}
\includegraphics[scale=0.5]{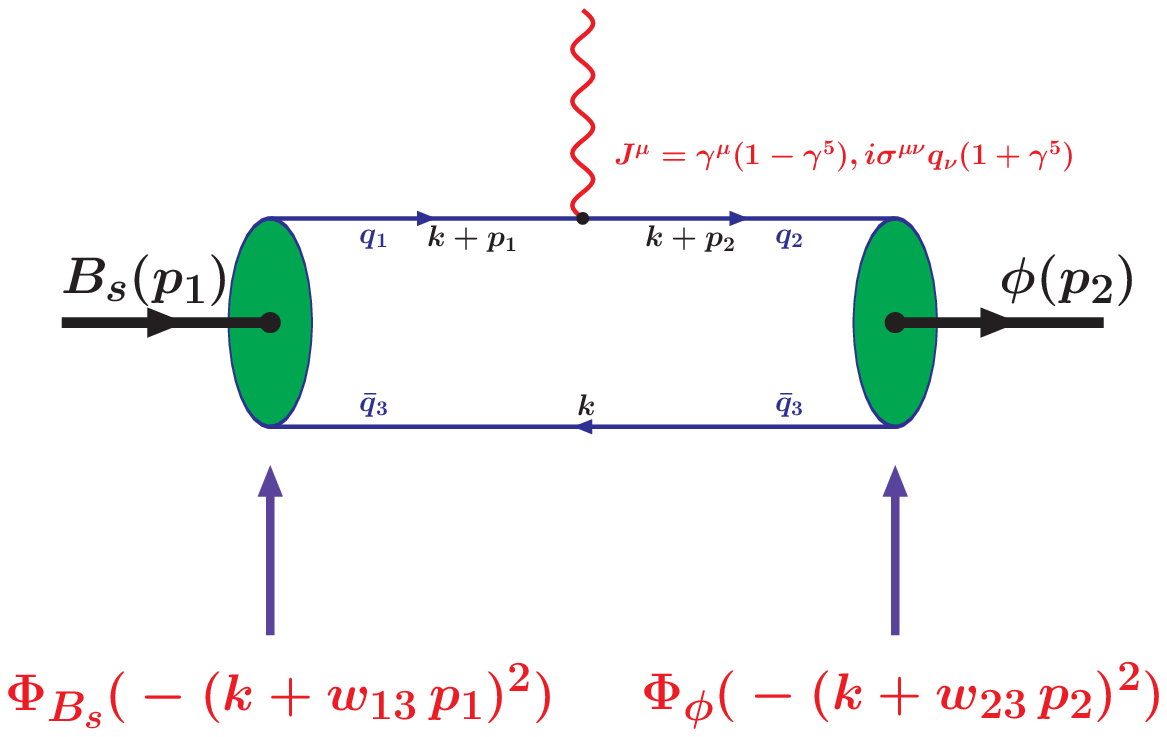}
\end{tabular}
\caption{A diagrammatic illustration of the matrix elements for $B_{s} \to \phi$ transitions is given, featuring the quark labeling $q_1=b$, $q_2=q_3=s$ along with the weight parameters $w_{12}=m_{s}/(m_{b}+m_{s})$ and $w_{23}=\frac{1}{2}$~\cite{Dubnicka:2016nyy}.}
\label{fig:Bs_phi}
\end{figure*}

The Feynman diagram illustrating the $B_{s} \to \phi$ transition within the covariant quark model is shown in Fig.~\ref{fig:Bs_phi}.

The matrix element can be written using a set of dimensionless form factors~\cite{Ivanov:2011aa,Dubnicka:2016nyy} as:
\bea
&&
\langle 
M_{2}(p_2,\epsilon_2)\,
|\,\bar s\, O^{\,\mu}\,b\, |\,M_{1}(p_1)
\rangle 
\,=\,
\nn
&=&
N_c\, g_{M_2}\,g_{M_1} \!\! \int\!\! \frac{d^4k}{ (2\pi)^4 i}\, 
\widetilde\Phi_{M_2}\Big(-(k+w_{13} p_1)^2\Big)\,
\widetilde\Phi_{M_{1}}\Big(-(k+w_{23} p_2)^2\Big)
\nn
&\times&
{\rm tr} \biggl[ 
O^{\,\mu} \,S_b(k+p_1)\,\gamma^5\, S_s(k) \not\!\epsilon_2^{\,\,\dagger} \,
S_s(k+p_2)\, \biggr]
\nn
 & = &
\frac{\epsilon^{\,\dagger}_{\,\nu}}{m_1+m_2}\,
\Big( - g^{\mu\nu}\,P\cdot q\,A_0(q^2) + P^{\,\mu}\,P^{\,\nu}\,A_+(q^2)
       + q^{\,\mu}\,P^{\,\nu}\,A_-(q^2)
\nn 
&& + i\,\varepsilon^{\mu\nu\alpha\beta}\,P_\alpha\,q_\beta\,V(q^2)\Big),
\label{eq:PV}
\ena
\bea
&&
\langle 
M_2(p_2,\epsilon_2)\,
|\,\bar s\, (\sigma^{\,\mu\nu}q_\nu(1+\gamma^5))\,b\, |\,M_1 (p_1)
\rangle 
\,=\,
\nn
&=&
N_c\, g_{M_2}\,g_{M_1} \!\! \int\!\! \frac{d^4k}{ (2\pi)^4 i}\, 
\widetilde\Phi_{M_2}\Big(-(k+w_{13} p_1)^2\Big)\,
\widetilde\Phi_{M_1}\Big(-(k+w_{23} p_2)^2\Big)
\nn
&\times&
{\rm tr} \biggl[ 
(\sigma^{\,\mu\nu}q_\nu(1+\gamma^5))
\,S_b(k+p_1)\,\gamma^5\, S_s(k) \not\!\epsilon_2^{\,\,\dagger} \,S_s(k+p_2)\, 
\biggr]
\nn
 & = &
\epsilon^{\,\dagger}_{\,\nu}\,
\Big( - (g^{\mu\nu}-q^{\,\mu}q^{\,\nu}/q^2)\,P\cdot q\,a_0(q^2) 
       + (P^{\,\mu}\,P^{\,\nu}-q^{\,\mu}\,P^{\,\nu}\,P\cdot q/q^2)\,a_+(q^2)
\nn
&&
+ i\,\varepsilon^{\mu\nu\alpha\beta}\,P_\alpha\,q_\beta\,g(q^2)\Big).
\label{eq:PVT}
\ena

Here $M_1$ -- $B_s$ and $M_2$ -- $\phi$ mesons. $P=p_1+p_2$, $q=p_1-p_2$, $\epsilon_2^\dagger\cdot p_2=0$, $p_1^2=m_1^2\equiv m^2_{B_s}$, $p_2^2=m_2^2\equiv m^2_\phi$ and the weak matrix $O^{\,\mu} = \gamma^{\,\mu}(1-\gamma^5)$.
$w_{ij}=m_{q_j}/(m_{q_i}+m_{q_j})$ $(i,j=1,2,3)$, and $w_{ij}+w_{ji}=1$. 

The free parameters in CCQM are the masses of constituent quarks $m_q$, the parameters $\Lambda_H$, which describe the effective sizes of hadrons and the parameter of infrared cutoff $\lambda$, universal for all processes, characterizing the confinement region. The numerical values of these parameters are determined by fitting well-established physical quantities and their experimental values and/or by lattice QCD results. Such quantities include the constants of lepton decays of hadrons, as well as some well-defined widths of radiation decays $P\to\gamma\gamma$ and $V\to P\gamma$.

The values of the free parameters of the model, which are given in the equations ~(\ref{eq:CQM-fitmas}) (in units of GeV), give the best fit with experimental values:

\begin{equation}
\begin{tabular}{ccccccc}
     $m_{u/d}$        &      $m_s$        &      $m_c$       &     $m_b$ & $\lambda$   & $\Lambda_{B_{s}}$ & $\Lambda{\phi}$
\\
\hline
  \ 0.241 \   &   \ 0.428 \   &   \ 1.672 \   &   \ 5.046 \   &  \ 0.181 \ & \ 2.049 \  & \ 0.883 \   
\end{tabular}
,
\label{eq:CQM-fitmas}
\end{equation}



%

Table~\ref{tab:ff-comparison4} presents the form factor values from CCQM~\cite{Dubnicka:2016nyy} as BSW form factors (see eqs. 23-24 in \cite{Dubnicka:2016nyy}), and compares them with other theoretical approaches. The superscript on the form factors is omitted when making comparisons with other approaches.

It should be noted that the free parameters in the model are determined by minimizing
the functional
$\displaystyle \chi^2 = \sum\limits_i\frac{(y_i^{\text{exp.}}-y_i^{\text{theory}})^2}{\sigma^2_i}$,
where $\sigma_i$--experimental error. If the value of $\sigma$ is too small,
then its value is assumed to be 10 $\%$, thus the errors of the fitted parameters
are also on the order of 10 $\%$. Thus, the theoretical error of the CCQM model is approximately about 10 $\%$ for form factor values and the order of 20$\%$ at the level of widths. 

\begin{table}
\caption{Comparison of form factors at maximum recoil $q^2=0$}
\label{tab:ff-comparison4}
\begin{tabular*}{\textwidth}{@{\extracolsep{\fill}}cccc@{}}
\hline
\hline
Model     &  $A_0(0)$ &$A_1(0)$ &$A_2(0)$ \\
\hline

CCQM\cite{Dubnicka:2016nyy}  & $0.28\pm 0.03$ & $0.27\pm 0.03$ & $0.27\pm 0.03$ \\
LCSR~\cite{Ball:2004rg}  & $0.474\pm0.037$ & 0.311$\pm$0.029 & 0.234$\pm$0.028 \\
LCSR~\cite{Wu:2006rd}&  &$0.271\pm0.014$&$0.212\pm0.011$\\
PQCD~\cite{Jin:2020qfp} &0.262 & 0.247 & 0.239 \\
CQM~\cite{Ivanov:2011aa}&  & 0.29 & 0.28 \\

RQM~\cite{Faustov:2013pca}  & $0.322\pm 0.016$ & $0.320\pm 0.016$  & $0.318\pm 0.016$\\

PQCD~\cite{Ali:2007ff} & $0.30\pm 0.05$ &$0.19\pm0.04$&\\

CQM~\cite{Melikhov:2000yu} & 0.42 & 0.34 &0.31\\

PQCD~\cite{Li:2009tx} & $0.31\pm 0.07$ & $0.18^{+0.06}_{-0.05}$ & $0.12\pm0.03$\\

LCUM~\cite{Lu:2007sg} & 0.279 &0.232 &0.210\\

\hline
\hline
  Model   & $V(0)$  &$T_1(0)$ &$T_3(0)$ \\
\hline

CCQM\cite{Dubnicka:2016nyy} & $0.31\pm 0.03$ & $0.27\pm 0.03$ & $0.18\pm 0.02 $\\
LCSR~\cite{Ball:2004rg} &0.434$\pm$0.035& 0.349$\pm$0.033& $0.175\pm0.018$\\
LCSR~\cite{Wu:2006rd} &$0.339\pm0.017$ &$0.299\pm0.016$& $0.191 \pm 0.010$\\
PQCD~\cite{Jin:2020qfp} & 0.311 & 0.264 & 0.196 \\
CQM~\cite{Ivanov:2011aa}& 0.32 & 0.28 & \\

RQM~\cite{Faustov:2013pca} & $0.406\pm 0.020$ & $0.275\pm0.014$& $0.133\pm 0.006$\\

PQCD~\cite{Ali:2007ff} &$0.25\pm0.05$&&\\

CQM~\cite{Melikhov:2000yu} &$0.44$  &0.38& 0.26\\

PQCD~\cite{Li:2009tx} &$0.26\pm0.07$& $0.23^{+0.06}_{-0.05}$& $0.19\pm 0.05$\\

LCUM~\cite{Lu:2007sg}  &0.329 &0.276& 0.170\\

\hline\hline
\end{tabular*}
\end{table}

\section{ Effective Hamiltonian }

We describe $b \to s \ell^+ \ell^-$ rare decay in terms of the
effective Hamiltonian~\cite{Buchalla:1995vs} as 
\be
{\mathcal H}_{\rm eff} = - \frac{4G_F}{\sqrt{2}} \lambda_t   
              \sum_{i=1}^{10} C_i(\mu)  \ord_i(\mu) ,
\label{eq:effHam}
\en
where $C_i(\mu)$ -- the Wilson coefficients and $\ord_i(\mu)$ -- local operators. $\lambda_t~=~|V_{tb}V_{ts}^\ast|$ is the product of the elements of the Cabibbo-Kobayashi-Maskawa matrix. It should be noted that small corrections proportional to $\lambda_u = |V_{ub}V_{us}^\ast|$ are discarded due to the relatively small values.
The standard set of local operators\cite{Buchalla:1995vs} obtained within the SM framework for~$b \to s l^+ l^-$ transition is written as
\bea 
\begin{array}{ll} 
\ord_1     =  (\bar{s}_{a_1}\gamma^\mu P_L c_{a_2})
              (\bar{c}_{a_2}\gamma_\mu P_L b_{a_1}),                   &
\ord_2     =  (\bar{s}\gamma^\mu P_L c)  (\bar{c}\gamma_\mu P_L b),   
\\[2ex]
\ord_3     =  (\bar{s}\gamma^\mu P_L  b) \sum_q(\bar{q}\gamma_\mu P_L q),  &
\ord_4     =  (\bar{s}_{a_1}\gamma^\mu P_L  b_{a_2}) 
              \sum_q (\bar{q}_{a_2}\gamma_\mu P_L q_{a_1}),
\\[2ex]
\ord_5     =  (\bar{s}\gamma^\mu P_L b)
              \sum_q(\bar{q}\gamma_\mu P_R q),            &
\ord_6     =  (\bar{s}_{a_1}\gamma^\mu P_L b_{a_2 }) 
              \sum_q  (\bar{q}_{a_2} \gamma_\mu P_R q_{a_1}),               
\\[2ex]
\ord_7     =  \frac{e}{16\pi^2} \bar m_b\, 
              (\bar{s} \sigma^{\mu\nu} P_R b) F_{\mu\nu},       &
\ord_8    =  \frac{g}{16\pi^2} \bar m_b\, 
              (\bar{s}_{a_1} \sigma^{\mu\nu} P_R {\bf T}_{a_1a_2} b_{a_2}) 
              {\bf G}_{\mu\nu},            
\\[2ex]
\ord_9     = \frac{e^2}{16\pi^2}        
             (\bar{s} \gamma^\mu P_L b) (\bar\ell\gamma_\mu \ell),     &
\ord_{10}  = \frac{e^2}{16\pi^2} 
             (\bar{s} \gamma^\mu P_L b)  (\bar\ell\gamma_\mu\gamma_5 \ell), 
\end{array}
\label{eq:operators}
\ena
where ${\bf G}_{\mu\nu}$ -- gluon field strength and $F_{\mu\nu}$ -- photon field strengths; ${\bf T}_{a_1a_2}$ are the generators of the $SU(3)$ color group; $a_1$ and $a_2$ denote the color indices, and are omitted in the color singlet currents.
The chirality projection operators are $P_{L,R} = (1 \mp \gamma_5)/2$, and $\mu$ is the renormalization scale.
$\ord_{1,2}$ -- current-current operators, $\ord_{3-6}$ -- QCD penguin operators, $\ord_{7,8}$ -- "magnetic penguin" operators, and $\ord_{9,10}$ -- the semileptonic electroweak penguin operators. The masses of QCD quarks are denoted by a bar to distinguish them from the masses of the constituent quarks used in CCQM.

The matrix element of the exclusive transition $B_s\to \phi \ell^+ \ell^-$ can be written by using the effective Hamiltonian defined by eq.(\ref{eq:effHam}) as 
\bea
{\mathcal M} & = & 
\frac{G_F}{\sqrt{2}}\cdot\frac{ \alpha\lambda_t}{\pi} \cdot
\Big\{
C_9^{\rm eff}\,<M_2\,|\,\bar{s}\,\gamma^\mu\, P_L\, b\,|\,M_1> 
\left( \bar \ell \gamma_\mu \ell \right)
\nn
&-& \frac{2\bar m_b}{q^2}\,C_7^{\rm eff}\, 
<M_2\,|\,  \bar{s}\,i\sigma^{\mu \nu} q_\nu \,P_R\, b\, |\,M_1>  
\left( \bar \ell \gamma_\mu \ell \right)
\nn
&+& C_{10}\, <M_2\,|\,\bar{s}\,\gamma^\mu P_L \, b\,|\,M_1>
\left(\bar \ell \gamma_\mu \gamma_5 \ell\right)
\Big\},
\label{eq:matrix-elem}
\ena
where $M_1$ -- $B_s$, $M_2$ -- $\phi$ mesons and $C_7^{\rm eff}= C_7 -C_5/3 -C_6$.
It should be noted that the matrix element in equation (\ref{eq:matrix-elem}) contains both the free quark decay amplitude coming from the operators $\ord_7$, $\ord_9$, and $\ord_{10}$ (gluon magnetic
penquin  $\ord_{8}$ does not contribute), and, in addition, certain long-range effects from the matrix elements of the four-quark operators $\ord_i\,\,(i=1,\ldots,6)$, which are usually absorbed by the redefinition of the Wilson coefficients at short distances.

The coefficient $C_9^{\rm eff}$ accounts for the contributions of the four-quark operators $\ord_i$ ($i=1,...,6$) and for nonperturbative $c\bar c$-resonance effects. These resonant contributions are parameterized, as is standard, by a Breit-Wigner ansatz \cite{Ali:1991is}.

The calculations in Refs.~\cite{Asatryan:2001zw,Greub:2008cy,Dubnicka:2016nyy} provide the two-loop contributions, which effectively renormalize the Wilson coefficients in addition to the perturbative charm-loop part
\bea
 C_7^{\rm eff} &\to&  C_7^{\rm eff}
             - \frac{\alpha_S}{4\pi}\Big( C_1 F_1^{(7)} + C_2 F_2^{(7)} \Big)\,,
\nn
 C_9^{\rm eff} &\to&  C_9^{\rm eff}
             - \frac{\alpha_S}{4\pi}\Big( C_1 F_1^{(9)} + C_2 F_2^{(9)} \Big)\,.
\label{eq:Greub}
\ena

The Standard Model Wilson coefficients are adopted from Ref.~\cite{Descotes-Genon:2013vna}. These coefficients were initially computed at the high-energy matching scale $\mu_0=2 M_W$ and subsequently evolved down to the hadronic scale $\mu_b= 4.8$~GeV, with the running of couplings and current quark masses treated consistently.

In the specific analysis of $b\to s\ell\ell$ transitions from \cite{Descotes-Genon:2015uva}, these inputs were used in a calculation incorporating Next-to-Next-to-Leading Logarithmic (NNLL) corrections. The impact of these higher-order logarithmic terms was found to be significant, shifting predictions by approximately 15~$\%$~\cite{Dubnicka:2016nyy}.

The SM Wilson coefficients are taken from the paper~\cite{Descotes-Genon:2013vna}.
They were calculated at the matching scale $\mu_0=2 M_W$ and reduced to the hadronic scale $\mu_b= 4.8$~GeV.
The evolution of the couplings and current quark masses occurs similarly.

The values of the model-independent input parameters and Wilson coefficients are given in the Table~\ref{tab:input}. 
\bgroup 
\def\arraystretch{1.3}
\begin{table}[htbp]
\caption{Input parameter values. $m_W$, $\bar m_c$, $\bar m_b$, $\bar m_t$ are given in GeV}
\vspace*{2mm}  
\centering
 \begin{tabular}{ccccccccc}
\hline\hline
  $m_W$ &  $\sin^2\theta_W $ &  $\alpha(M_Z)$ & 
$\bar m_c$ &  $\bar m_b$  &  $\bar m_t$ & $ \lambda_t$  & &\\
\hline
 $80.41$ & $0.2313$ & $1/128.94$ & $1.27$ & $4.68$ & $173.3$ &  
 0.041 & & \\
\hline\hline
\quad $C_1$ \quad & \quad $C_2$ \quad & \quad $C_3$ \quad & \quad $C_4$ \quad & \quad $C_5$ \quad & \quad $C_6$ \quad & \quad $C^{\rm eff}_7$ \quad & \quad $C_9$ \quad & \quad $C_{10}$ \quad \\
\hline
 $-0.2632$ & $1.0111$ &  $-0.0055$ & $-0.0806$ & 0.0004 & 0.0009 & $-0.2923$ &
  4.0749 & $-4.3085$ \\
\hline\hline
 \end{tabular}
\label{tab:input}
\end{table}
\egroup
\section{Numerical results}
The meson masses and the $B_s$-meson’s lifetime was taken from the PDG~\cite{ParticleDataGroup:2024cfk} and as follows $m_{B_{s}}=5366.3$ MeV, $m_{\phi}=1019.455$ MeV, $m_{J/\psi}=3096.916$ MeV and $\tau_{B_{s}}=1.51 \times 10^{-12} s$.

The fourfold angular distribution in the cascade decay  $B\to\phi(\to K^+K^-)\bar \ell \ell$ provides access to numerous physical observables. For the $B_s\to\phi\mu^+\mu^-$ decay
\cite{Aaij:2015esa,LHCb:2021zwz,LHCb:2021xxq}, these include the CP-averaged differential branching ratio
$d{\cal B}/dq^2$, the CP-averaged longitudinal polarization fraction $F_L$, the forward-backward asymmetry $A_{FB}$
and angular observables $S_{3,4,7}$. The latter are related to the optimized observables $P_i$
\cite{Descotes-Genon:2015uva}.
In the Standard Model, the CP asymmetries corresponding to the observables $A_{5,6,8,9}$ \cite{Bobeth:2008ij} arise from the weak phase in the CKM matrix. Specifically, for  $b\to s$ transitions, these asymmetries are proportional to
${\rm Im}(\hat\lambda_u)\equiv {\rm Im}(V_{ub}V^\ast_{us}/V_{tb}V^\ast_{ts})$, a parameter of order $10^{-2}$~\cite{Bobeth:2008ij}. 

The width of $B_s\to \phi\bar \ell \ell$ decay is calculated by integrating the $q^2$-differential distribution
\bea
&&
\frac{d\Gamma(B\to \phi\bar \ell \ell)}{dq^2} =\,
 \frac{G^2_F}{(2\pi)^3}\,
\left(\frac{\alpha \lambda_t}{2\pi}\right)^2
\frac{|{\bf p_2}|\,q^2\,\beta_\ell}{12\,m_{B_s}^2}
{\cal H}_{\rm tot}\,,
\nn[1.2ex]
&&
{\cal H}_{\rm tot} = 
\frac12\left(  {\cal H}^{11}_U + {\cal H}^{22}_U 
             + {\cal H}^{11}_L + {\cal H}^{22}_L \right )
+ \delta_{\ell\ell}\,
\left[\,\frac12 {\cal H}^{11}_U - {\cal H}^{22}_U
       + \frac12 {\cal H}^{11}_L - {\cal H}^{22}_L + \frac32\, {\cal H}^{22}_S
\right],
\label{eq:distr1}
\ena
where $\beta_\ell=\sqrt{1-4m_\ell^2/q^2}$ and $\delta_{\ell\ell} = 2m^2_\ell/q^2$.
$|{\bf p_2}|=\lambda^{1/2}(m_1^2,m_2^2,q^2)/(2\,m_1)$ -- the momentum of the $\phi$-meson, specified in the rest frame of $B_s$ meson.
Combinations of helicity amplitudes ${\cal H}$ are represented as follows~(see the papers~\cite{Faessler:2002ut,Dubnicka:2016nyy} for details) :
\bea
{\cal H}^{ii}_U   &=&  |H^i_{+1 +1}|^2 +  |H^i_{-1 -1}|^2, \qquad
{\cal H}^{ii}_L   = |H^i_{00}|^2, \qquad 
{\cal H}^{ii}_S   = |H^i_{t0}|^2.
\label{eq:bilinear}
\ena

The differential rate of the $B_s\to\phi\nu\bar\nu$ decay can be found as follows 

\be
\frac{d\Gamma(B_s\to\phi\nu\bar\nu)}{dq^2} 
= \frac{G_F^2}{(2\pi)^3} \Big(\frac{\alpha\lambda_t}{2\pi}\Big)^2
\Big[\frac{D_\nu(x_t)}{\sin^2\theta_W}\Big]^2
\frac{|{\bf p_2}|\, q^2}{4m_1^2}\cdot (H_U+H_L)\,,
\en
where $x_t=\bar m_t^2/m_W^2$ and the function $D_\nu$ is given by
\be
D_\nu(x) = \frac{x}{8}\left(\frac{2+x}{x-1}+\frac{3x-6}{(x-1)^2}\,\ln x\right).
\en
The corresponding combinations of bilinear helicities are defined as
\bea
{\cal H}_U   &=&  |H_{+1 +1}|^2 +  |H_{-1 -1}|^2, \qquad 
{\cal H}_L   = |H_{00}|^2, 
\nn[1.2ex]
H_{\pm1\pm1} &=& 
\frac{1}{m_1+m_2}\left[(m_{2}^{2}-m_{1}^{2})\, A_0\pm 2\,m_1\,|{\bf p_2}|\, V \right],
\nn[1.2ex]
H_{00} &=&  
\frac{1}{m_1+m_2}\frac{1}{2\,m_2\sqrt{q^2}} 
\left[(m_{2}^{2}-m_{1}^{2})\,(m_1^2 - m_2^2 - q^2)\, A_0 + 4\,m_1^2\,|{\bf p_2}|^2\, A_+\right].
\label{eq:bilinear-2}
\ena

The width of the color-suppressed nonleptonic decay
$B_s\to J/\psi\phi$ is determined by the formula\cite{Ivanov:2011aa}
\bea
\Gamma(B_s\to J/\psi\phi ) &=&
\frac{G_F^2}{16\pi}\frac{|{\bf p_{\,2}}|}{m^2_{1}}  
|V_{cb}V_{cs}|^2 
\left(C^{\,\rm eff}_1+ C^{\,\rm eff}_5\right)^2 
\left( m_{J/\psi}\,f_{J/\psi} \right)^2\,(H_U+H_L) ,
\label{eq:BsJpsiPhi}
\ena
where the square of the transferred momentum is taken from the mass $J/\psi$, that is
$q^2=m^2_{J/\psi}$, $V_{cb}~=~0.406$, $V_{cs}=0.975$ and $f_{J/\psi}=415$~MeV. The Wilson coefficients are equal to
$ C^{\,\rm eff}_{1}=C_1+\xi\, C_2+C_3+\xi\, C_4 $ and
$ C^{\,\rm eff}_{5}=C_5+\xi\, C_6$ according to the naive factorization.
In accordance with the $1/N_c$ expansion, terms scaled by the color factor $\xi = 1/N_c$ are omitted from the numerical evaluation.

The radiative decay width for the process $B_s\to\phi\gamma $ is then given by the following expression
\be
\Gamma(B_s\to\phi\gamma)
= \frac{G_F^2\alpha\lambda_t^2}{32\pi^4}
\bar m_b^2 m_1^3\Big(1-\frac{m_2^2}{m_1^2}\Big)^3\,|C^{\rm eff}_7|^2\, g^2(0)\,.
\en

Table~\ref{tab:branching} presents theoretical predictions and experimental data~\cite{ParticleDataGroup:2024cfk,Aaij:2015esa,LHCb:2021zwz} for branching decays $B_s\to \phi\mu^+\mu^- $,  $B_s\to  \phi\tau^+\tau^- $, ${\cal B}(B_s\to\phi \nu\bar\nu)$, $B_s\to  \phi\gamma $ and $B_s\to \phi J/ \psi$ .

\begin{table}[ht]
\caption{Comparison of CCQM predictions~\cite{Dubnicka:2016nyy} for total branchings with experimental data and predictions from other theoretical approaches}
\label{tab:branching}
\begin{tabular*}{\textwidth}{@{\extracolsep{\fill}}c|cccccc@{}}
\hline
  \hline
 & CCQM~\cite{Dubnicka:2016nyy}   & LFQM~\cite{Geng:2003su} & PQCD~\cite{Jin:2020qfp} & RQM~\cite{Faustov:2013pca}
 &LHCb~\cite{Aaij:2015esa,ParticleDataGroup:2024cfk} & LHCb~\cite{LHCb:2021zwz} \\
\hline
 $10^7 {\cal B}(B_s\to\phi \mu^+\mu^-)$ & $9.11\pm 1.82$ 
 & 16.4 & $7.07\pm 2.81$ & $11.1\pm 1.1$ & $ 7.97 \pm 0.77 $ & $8.14 \pm 0.472$\\
\hline 
$10^7 {\cal B}(B_s\to\phi \tau^+\tau^-)$  & $1.03\pm 0.20$ 
 & 1.60 & $0.81\pm 0.35$ & $1.03\pm 0.20$& &\\
 \hline 
 $10^2 {\cal B}(B_s\to J/\psi \phi)$ & $0.16\pm 0.03$ 
&&&  $0.113\pm 0.016$& $0.108\pm 0.009$ &$0.102\pm 0.010$ \\
\hline  
 $10^5 {\cal B}(B_s\to\phi \nu\bar\nu) $   & $0.84\pm 0.16$ 
 & 1.165 & &$0.796\pm 0.080 $& $< 540 $ &  \\
  \hline 
$10^5 {\cal B}(B_s\to\phi\gamma) $ & $ 2.39\pm 0.48$ 
 & && $3.8\pm 0.4$ & $3.52\pm 0.34$& \\
\hline\hline
\end{tabular*}
\end{table}

As detailed in \cite{Faessler:2002ut}, the full angular decay distribution for this rare $B$ decay is formulated using helicity amplitudes and incorporates lepton mass effects. 
This distribution, which depends on three angles and the invariant mass $q^2$ of the lepton pair, enables the definition of multiple physical observables that are experimentally accessible.
The observables in this analysis include the branching ratio, the forward-backward asymmetry longitudinal polarization fraction of the $\phi$ meson.
The branching ratio is determined by integrating the complete fourfold angular decay distribution over the entire angular phase space. 
This results in the explicit expression provided in equation (\ref{eq:distr1}), which is formulated using helicity amplitudes.
The derivation of the required ratio between these helicity amplitudes and the transverse amplitudes is given in \cite{Dubnicka:2016nyy}.
Using the helicity amplitudes, one can write $F_L$ and $A_{FB}$ as follows:
\bea
F_L &=&  
\frac12 \beta_\ell^2 
\frac{ {\cal H}_L^{11} +  {\cal H}_L^{22}}{ {\cal H}_{\rm tot} }, 
\label{eq:FL}\\[2ex]
A_{\rm FB} &=& 
\frac{1}{d\Gamma/dq^2} \left[ \int\limits_0^1 - \int\limits_{-1}^0 \right]
d\!\cos\theta\, \frac{d^2\Gamma}{dq^2 d\!\cos\theta} 
= -\frac34\beta_\ell \frac{ {\cal H}_P^{12}} { {\cal H}_{\rm tot} }\,,
\label{eq:AFB}
\ena
where $\theta$ denotes the polar angle between the $\ell^+\ell^-$ plane and the $z$-axis. As these quantities are defined as ratios of hadronic amplitudes, $A_{\rm FB}$ and $F_L$ are expected to exhibit a reduced dependence on theoretical uncertainties.

\begin{figure*}[htbp]
\centering
\begin{tabular}{lr}
\includegraphics[scale=0.6]{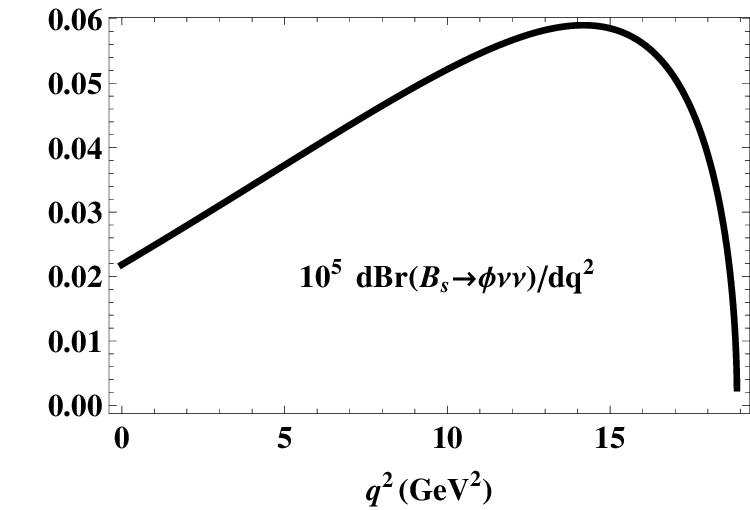}
\end{tabular}
\caption{The behavior of the $ {\cal B}(B_{s} \to \phi \nu \bar\nu)$ depending on the transferred momentum $q^2$} 
\label{fig:Br_nu}
\end{figure*}

The behavior of the differential branching ${\cal B}(B_s\to\phi \nu\bar\nu)$ is shown in Fig.~\ref{fig:Br_nu}

To minimize hadronic uncertainties, a set of optimized observables $P_i$ was constructed by taking appropriate ratios of form factors \cite{Descotes-Genon:2013vna}. This construction aimed to reduce form factor dependence, maximize the sensitivity to New Physics beyond the SM, and ensure the observables could be measured experimentally. Despite these advantages, they are more difficult to assign a clear physical meaning, in contrast to observables such as $A_{FB}$ and $F_L$.

\begin{figure*}[htbp]
\centering
\begin{tabular}{lr}
\includegraphics[scale=0.6]{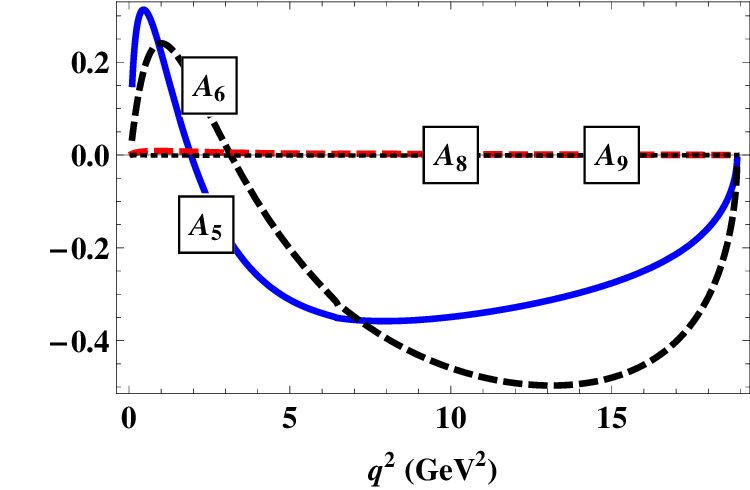} &
\includegraphics[scale=0.6]{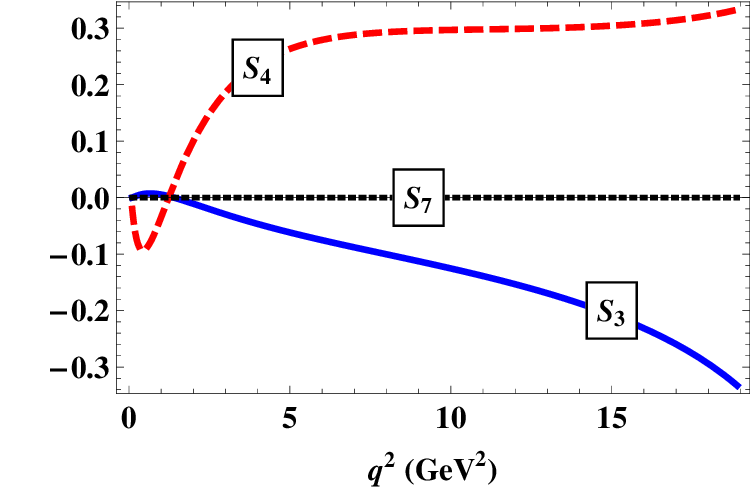}\\
\end{tabular}
\caption{Observables $A_{i}$ and $S_{i}$ for $B_s\to\phi \mu^+\mu^-$ decay}
\label{fig:P14}
\end{figure*}

\begin{figure*}[htbp]
\centering
\begin{tabular}{lr}
\includegraphics[scale=0.6]{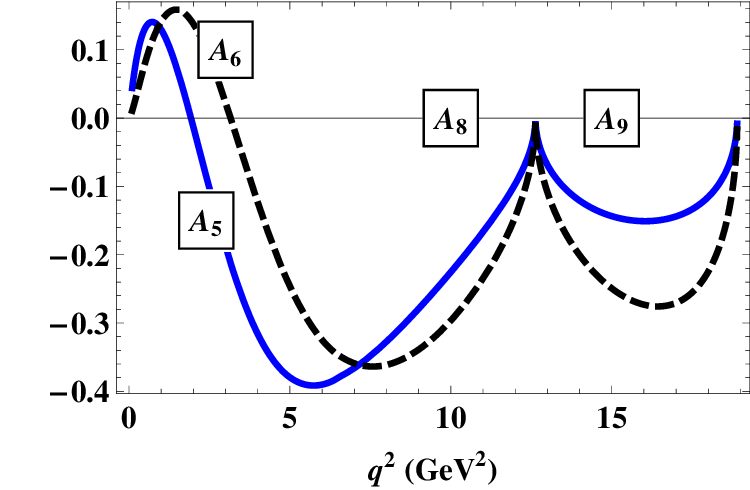} &
\includegraphics[scale=0.6]{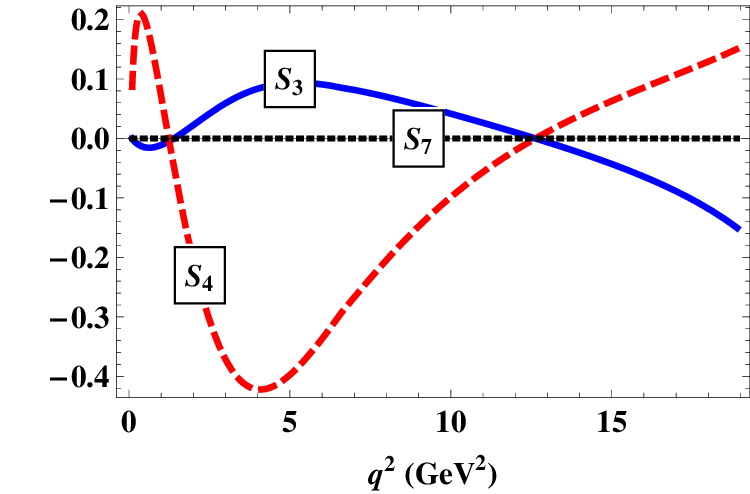}\\
\end{tabular}
\caption{Observables $A_{i}$ and  $S_{i}$ for $B_s\to\phi \tau^+\tau^-$ decay}
\label{fig:P15}
\end{figure*}

The optimized observables are not stated in \cite{Aaij:2015esa}. The corresponding numerical values for them were obtained in \cite{Descotes-Genon:2015uva} through a conversion of the results given for $S_{3,4,7}$.
Calculations of $S_{3,4,7}$ were performed in the CCQM, following the relations established in \cite{Descotes-Genon:2012isb}:
\be
S_3 = \frac12 F_T P_1, \qquad 
S_4 = \frac12 \sqrt{F_T F_L} P'_4, \qquad
S_7 = - \sqrt{F_T F_L} P'_6 .
\en
The $q^2$-dependence of the $A_{i}$ and $S_{i}$ for the decays $B_s\to\phi \mu^+\mu^-$ and $B_s\to\phi \tau^+\tau^-$ is shown in Fig.\ref{fig:P14} and Fig.\ref{fig:P15}, respectively.

The average values of the angular observables, integrated over the full $q^2$ range, are shown in Table~\ref{tab:obs-numerics}.
The table compares our results from the CCQM with the PQCD predictions~\cite{Jin:2020qfp}.

\begin{table*}[ht] 
\begin{center}
\caption{Comparison of $q^2$ -- averaged angular observables between CCQM and PQCD~\cite{Jin:2020qfp}}
\label{tab:obs-numerics}
\begin{tabular*}{\textwidth}{@{\extracolsep{\fill}}c|cccccc@{}}
\multicolumn{7}{c}{ $B_s\to\phi \mu^+\mu^-$ }\\
\hline
& $<A_{FB}>$       
& $<F_L>$ 
& $<P_1>$   
& $<P_4'>$ 
& $<S_3>$ 
& $<S_4>$
\\
\hline
 CCQM    
& $ -0.23\pm 0.05 $  
& $  0.46\pm 0.09 $  
& $ -0.52\pm 0.1$   
& $  1.04\pm 0.21$   
&$ -0.14\pm 0.03$
& $ 0.27\pm 0.05$   
\\
PQCD~\cite{Jin:2020qfp}     
& $ -0.233\pm 0.004$   
& $  0.454 ^{+0.006}_{-0.007}$  
& $ -0.555 ^{+0.010}_{-0.012} $    
& $ 1.111 ^{+0.011}_{-0.012}$   
& $ -0.144\pm 0.005$
&  $ 0.258^{+0.003}_{-0.004}$      
\\
\hline
\hline
\multicolumn{7}{c}{ $B_s\to\phi \tau^+\tau^-$ }\\
\hline
& $<A_{FB}>$       
& $<F_L>$ 
& $<P_1>$   
& $<P_4'>$ 
& $<S_3>$ 
& $<S_4>$
\\
\hline
 CCQM   
& $ -0.18\pm 0.04$   
& $  0.090\pm 0.02$  
& $ -0.76\pm 0.15$    
& $ 1.33\pm 0.27$   
& $ -0.067\pm 0.013$
&  $ 0.083\pm 0.017$
\\
PQCD~\cite{Jin:2020qfp}   
& $ -0.171\pm 0.002$   
& $  0.396^{+0.002}_{-0.003}$  
& $ -0.795^{+0.007}_{-0.004}$    
& $ 1.33\pm 0.002$   
& $ -0.080\pm 0.001$
&  $ 0.100\pm 0.001$      
\\
\hline
\hline
\end{tabular*}
\end{center}
\end{table*}

The optimized observable $S_7$ is precisely zero. This follows from the use of real form factors in conjunction with NLL-order Wilson coefficients, where an imaginary part is present only in $C_9^{\rm eff}$.
Furthermore, the optimized observable $P_1$ remains small across a large range of values for any choice of Wilson coefficients.
It is easy to verify that $P_1\propto A_0(0)-V(0)$ for $q^2=0$. 
In CCQM, the values $V(0)=0.31$ and $A_0(0)=0.40$ deviate from the heavy quark limit equality, producing a very small $P_1$.

Further, the calculation of two-loop corrections for the $b\to s\ell^+\ell^-$ decay has been addressed in distinct $q^2$ regions. For the low-$q^2$ regime, \cite{Asatryan:2001zw} provided results as an expansion in the small parameters $\hat s=q^2/\bar m_b^2$ and $z=\bar m_c^2/\bar m_b^2$, considering the range $0.05 \leq \hat s \leq 0.25$. The NNLO corrections for the high-$q^2$ region above the charm threshold ($q^2 > 4\bar m_c^2$) were derived in \cite{Greub:2008cy}, where the region was limited to $0.4 \le \hat s \le 1.0$.
The valid $q^2$ regions for the two-loop corrections are determined using the QCD bottom quark mass $\bar m_b = 4.68$ GeV from Table~\ref{tab:input}:
\be
1.1 \le q^2 \le 5.5 \, \, \text{GeV}^2 \quad \text{(small region)}
\quad
\text{and}
\quad
8.8 \le q^2 \le 22 \, \, \text{GeV}^2 \quad \text{(large region)}.
\label{eq:NNLL}
\en
The two-loop calculation for low $q^2$ \cite{Asatryan:2001zw} is valid under the conditions that $q^2/\bar m_b^2 \ll 1$ and $q^2/(4\bar m_c^2) \ll 1$, making its results reliable for $0.1 \le q^2 \le 6$ GeV$^2$. In contrast, the high-$q^2$ expansion \cite{Greub:2008cy} requires $q^2/\bar m_b^2 > 0.4$. 
We therefore omit the intervals $[5,8]$ and $[6,8]$ GeV$^2$ from the two-loop consideration due to this constraint.

The inclusion of NNLL corrections results in non-zero values for the observables $P'_6$ and $S_7$. These corrections are significant, contributing up to 20$\%$ at low squared momentum transfer ($q^2 \le 6$~GeV$^2$), but their impact becomes truly negligible at high $q^2$~\cite{Dubnicka:2016nyy}.

The CCQM results for the branching of ${\cal B}(B_s\to \phi\mu^+\mu^-)$ decay and angular observables are summarized in Table~\ref{tab:bin1}. The calculations include NNLL corrections to the Wilson coefficients \cite{Asatryan:2001zw,Greub:2008cy}, leading to effective modifications of $C_7^{\rm eff}$ and $C_9^{\rm eff}$. Thus, all values are derived from two-loop calculations with the exception of the $[5,8]$ and $[6,8]$ GeV$^2$ bins (which use one-loop calculations).

\begin{longtable}{@{}ccccc@{}}
\caption{CCQM predictions for branching and angular observables of ${\cal B}(B_s\to \phi\mu^+\mu^-)$ decay by bins compared with LHCb data and values from PQCD~\cite{Jin:2020qfp}}\\
\label{tab:bin1}
\endfirsthead
\endhead
\toprule[1.2pt]
\hline\hline
  $10^7 {\cal B}$    &   CCQM     &    PQCD~\cite{Jin:2020qfp}     &    LHCb~\cite{Aaij:2015esa}   &  LHCb~\cite{LHCb:2021zwz} \\ 
$[0.1,0.98]$ & $0.66\pm 0.13$  & $0.25^{+0.10}_{-0.07}$  &   & $0.68 \pm 0.06$ \\

 $[0.1,2]$ & $0.99\pm 0.2$  & $0.40^{+0.16}_{-0.11}$  & $1.11 \pm 0.16$   & \\ 
 $[1.1,2.5]$  & $0.42\pm 0.08$  & $0.21^{+0.07}_{-0.05}$ &  & $0.44 \pm 0.05$ \\
 $ [1,6]$ & $1.56\pm 0.31$ &$1.10^{+0.34}_{-0.25}$	 & $ 1.29 \pm 0.19 $  & \\
 $[1.1,6]$ & $1.52\pm 0.30$   & $0.98^{+0.30}_{-0.22}$&  & $1.41 \pm 0.10$ \\
  $[2.5,4]$ & $0.46\pm 0.09$   & $0.27^{+0.09}_{-0.06}$ 
&  & $0.35 \pm 0.04$ \\
 $[2,5]$   & $0.90\pm 0.18$  & $0.58^{+0.18}_{-0.13}$  & $0.77\pm 0.14$ & \\ 
 $[4,6]$   & $0.65\pm 0.13$  & $0.50^{+0.14}_{-0.11}$ & & $0.62\pm 0.06$ \\ 
 $[5,8]$    & $1.25\pm 0.25$
         & $0.96^{+0.25}_{-0.19}$  & $0.96\pm 0.15$ &  \\ 
 $[6,8]$    & $0.87\pm 0.17$
         & $0.69^{+0.18}_{-0.13}$  & &$0.63\pm 0.06$   \\
 $ [11,12.5]$ & $0.84\pm 0.17$ & $0.76^{+0.14}_{-0.11}$  & $ 0.71 \pm 0.12 $ & $ 0.72 \pm 0.06 $ \\ 
 $ [15,17]$ & $1.15\pm 0.23$ &$0.99^{+0.11}_{-0.10}$
 				 & $ 0.90 \pm 0.13 $ & $ 1.05 \pm 0.08 $\\ 
 $ [17,18.9]$ & $0.75\pm 0.15$ &$0.67^{+0.05}_{-0.05}$
& $ 0.75 \pm 0.13 $ & $ 0.82 \pm 0.07 $ \\ 

$[15,18.9]$  &  $1.89\pm 0.28$    &  $1.60^{+0.16}_{-0.16}$ 
           &  $1.62\pm 0.20$    &  $1.85\pm 0.13$  \\ 
\hline

$ F_L$  & CCQM & PQCD~\cite{Jin:2020qfp}  & LHCb~\cite{Aaij:2015esa} & LHCb~\cite{LHCb:2021xxq}\\ 
$[0.1,0.98]$ & $0.23\pm 0.05$ & 
         && $ 0.254 \pm 0.048 $ \\
 $[0.1,2]$ & $0.37\pm 0.07$  &  $0.472^{+0.011}_{-0.012}$   
          & $0.20\pm 0.09$&   \\ 
$[1.1,4]$ & $0.72\pm 0.14$ &&  
& $ 0.723 \pm 0.055 $ \\

 $[2,5]$   & $0.72\pm 0.14$  & $0.796^{+0.007}_{-0.007}$
  & $0.68\pm 0.15$ &  \\

$[4,6]$ & $0.65\pm 0.13$ & 
         & & $ 0.701 \pm 0.05 $ \\
 $[5,8]$   & $0.57 \pm 0.11$ & $0.682^{+0.010}_{-0.008}$ 
& $0.54\pm 0.10$  & $0.54\pm 0.10$   \\

$[6,8]$ & $0.55\pm 0.11$ &  
         & &$ 0.624 \pm 0.052 $ \\
  $ [11,12.5] $ & $0.40\pm 0.08$ & $0.524^{+0.008}_{-0.008}$
  				& $0.29 \pm 0.11$ & $ 0.35 \pm 0.046 $  \\ 
 $ [15,17] $ 	& $0.34\pm 0.07$ &$0.412^{+0.004}_{-0.004}$
 				& $0.23 \pm 0.096$ & \\ 
 $ [17,18.9] $ 	& $0.33\pm 0.06$ &$0.363^{+0.001}_{-0.001}$
 				& $0.40 \pm 0.15$ & $ 0.4 \pm 0.14 $ \\ 
 $ [1,6] $ 	& $0.69\pm 0.14$ &$0.777^{+0.008}_{-0.006}$
 				& $0.63 \pm 0.09$ &   \\ 
 $[1.1,6]$ & $0.69\pm 0.14$   & 
&  & $0.715 \pm 0.04$ \\

 $[15,18.9]$ & $0.34\pm 0.07$  & $0.394^{+0.003}_{-0.003}$
         & $0.29\pm 0.07$  & $0.35\pm 0.031$ \\ 
 %
\hline
$ A_{FB} $  & CCQM & PQCD~\cite{Jin:2020qfp} 
 &LHCb~\cite{Aaij:2015esa} & LHCb~\cite{LHCb:2021xxq}\\ 
$[0.1,0.98]$ & $0.11\pm 0.02$ & 
         && $ 0.068 \pm 0.0065 $ \\
 $[0.1,2]$ & $0.14\pm 0.02$  &  $0.122^{+0.003}_{-0.002}$   
          & &   \\ 
$[1.1,4]$ & $0.12\pm 0.02$ &&  
& $ 0.023 \pm 0.054 $ \\

 $[2,5]$   & $0.038\pm 0.008$  & $-0.038^{+0.001}_{-0.002}$
  &  &  \\

$[4,6]$ & $-0.085\pm 0.017$ &
         & & $-0.030 \pm 0.051 $ \\
 $[5,8]$   & $-0.24 \pm 0.05$ & $-0.200^{+0.006}_{-0.006}$ 
&   &    \\

$[6,8]$ & $-0.26\pm 0.05$ &  
         & &$ 0.32 \pm 0.049 $ \\
  $ [11,12.5] $ & $-0.35\pm 0.07$ & $-0.298^{+0.005}_{-0.005}$
  				&  & $ 0.034 \pm 0.048 $  \\ 
 $ [15,17] $ 	& $-0.33\pm 0.06$ &$-0.290^{+0.005}_{-0.005}$
 				&  & \\ 
 $ [17,18.9] $ 	& $-0.24\pm 0.05$ &$-0.208^{+0.004}_{-0.003}$
 				& &  \\ 
 $ [1,6] $ 	& $0.69\pm 0.14$ &$0.777^{+0.008}_{-0.006}$
 				&  &   \\ 
 $[1.1,6]$ & $0.034\pm 0.006$   & 
&  & $0.006 \pm 0.036$ \\

 $[15,18.9]$ & $-0.29\pm 0.06$  & $0.394^{+0.003}_{-0.003}$
         &  & $-0.011\pm 0.033$ \\ 
 %
\hline
$ S_3 $ & CCQM & PQCD~\cite{Jin:2020qfp}
 & LHCb~\cite{Aaij:2015esa} & LHCb~\cite{LHCb:2021xxq}\\ 
$[0.1,0.98]$ & $0.004\pm 0.001$ & 
         && $ -0.004 \pm 0.069 $ \\
 $[0.1,2]$ & $0.0031\pm 0.0006$  & $0.002^{+0.000}_{-0.001}$
             & $-0.05 \pm 0.13  $ & \\ 
$[1.1,4]$ & $-0.017\pm 0.003$ &&  
& $ -0.030 \pm 0.057 $ \\

 $[2,5]$   & $-0.035\pm 0.007$   & $-0.021^{+0.001}_{-0.001}$ 
         & $-0.06\pm 0.021$  &   \\ 

$[4,6]$ & $-0.059\pm 0.001$ &
& & $-0.162 \pm 0.068 $ \\

 $[5,8]$ & $-0.082\pm 0.016$   & $-0.050^{+0.004}_{-0.003}$
         & $-0.10\pm 0.27$     & $-0.10 \pm 0.25 $   \\ 
 $[6,8]$ & $-0.088\pm 0.016$   & 
&     & $0.013 \pm 0.081 $   \\

 $ [11,12.5] $ & $-0.15\pm 0.03$ & $-0.115^{+0.002}_{-0.001}$
  				 & $ -0.19 \pm 0.21 $ &$ -0.138 \pm 0.072 $ \\ 
 $ [15,17] $ & $-0.23\pm 0.05$  & $-0.213^{+0.001}_{-0.001}$
 				 & $ -0.06 \pm 0.18 $ &\\ 
 $ [17,18.9] $ & $-0.29\pm 0.06$  & $-0.281^{+0.001}_{-0.001}$
 				& $ -0.07 \pm 0.25 $& \\ 
 $ [1,6] $ & $ -0.034\pm 0.007$ & $-0.023^{+0.001}_{-0.001}$
& $ -0.02 \pm 0.13 $& \\

 $[1.1,6]$ & $-0.035\pm 0.007$   & 
&  & $-0.083 \pm 0.048$ \\
 $[15,18.9]$ & $-0.25\pm 0.05$     & $-0.239^{+0.001}_{-0.001}$
         & $-0.09\pm 0.12$     & $-0.247\pm 0.044$     \\ 
 %
\hline
$ S_4 $  & CCQM & PQCD~\cite{Jin:2020qfp} 
 & LHCb~\cite{Aaij:2015esa}& LHCb~\cite{LHCb:2021xxq}\\ 
$[0.1,0.98]$ & $-0.062\pm 0.012$ & 
&& $-0.213 \pm 0.082 $ \\

 $[0.1,2]$ & $-0.038\pm 0.008$   & $-0.053^{+0.000}_{-0.000}$
            & $-0.27 \pm 0.23 $ &\\ 
$[1.1,4]$ & $0.114\pm 0.023$ &&  
& $ 0.110 \pm 0.079 $ \\
 $[2,5]$   & $0.19\pm 0.04$      & $0.177^{+0.003}_{-0.003}$
               & $0.47\pm 0.37$&  \\ 
$[4,6]$ & $0.253\pm 0.051$ &
& & $0.222 \pm 0.092 $ \\

 $[5,8]$  & $0.28\pm 0.06$   & $0.259^{+0.003}_{-0.003}$
            & $0.10\pm 0.17$ & \\ 
 $[6,8]$ & $0.284\pm 0.057$   & 
&     & $0.176 \pm 0.078 $   \\

 $ [11,12.5] $ & $0.30\pm 0.06$ & $0.303^{+0.001}_{-0.001}$
 				 & $ 0.47 \pm 0.25 $ &$ 0.319 \pm 0.061 $ \\ 
 $ [15,17] $ & $0.31\pm 0.06$ & $0.323^{+0.001}_{-0.001}$
 				 & $ 0.03 \pm 0.15 $&\\ 
 $ [17,18.9] $ & $0.32\pm 0.06$ & $0.329^{+0.001}_{-0.001}$
 				 & $ 0.39 \pm 0.3 $&\\ 
 $ [1,6] $ & $0.17\pm 0.03$ & $0.169^{+0.001}_{-0.001}$
 				 & $ 0.19 \pm 0.14 $& \\ 

 $[1.1,6]$ & $0.174\pm 0.03$   & 
&  & $0.155 \pm 0.058$ \\

 $[15,18.9]$ & $0.31\pm 0.06$      & $0.325^{+0.001}_{-0.001}$
              & $0.14\pm 0.11$ &$0.208\pm 0.047$ \\ 
 %
\hline
$ S_7 $ & CCQM & PQCD~\cite{Jin:2020qfp}
 & LHCb~\cite{Aaij:2015esa} & LHCb~\cite{LHCb:2021xxq}\\ 
$[0.1,0.98]$ & $0.0056\pm 0.011$ & 
&& $-0.178 \pm 0.072 $ \\

 $[0.1,2]$ & $0.0065\pm 0.0013$  & $1.551^{+0.003}_{-0.006}$
           &  $0.04\pm 0.12$  & \\ 
$[1.1,4]$ & $0.0075\pm 0.0015$ &&  
& $ -0.101 \pm 0.075 $ \\

 $[2,5]$   & $0.0065\pm 0.0013$  &  $0.979^{+0.017}_{-0.018}$
           &  $-0.03\pm 0.21$ & \\ 
$[4,6]$ & $0.0051\pm 0.0010$ &
& & $0.175 \pm 0.089 $ \\

 $[5,8]$   &   &  $0.453^{+0.003}_{-0.003}$
&  $0.04\pm 0.18$ &\\
$[6,8]$   &  &  $0.453^{+0.003}_{-0.003}$
            &  & $0.033\pm 0.081$\\ 
 $ [11,12.5] $ & $ 0.0021\pm 0.0004$ & $0.185^{+0.001}_{-0.001}$
 			 & $ 0.00 \pm 0.16 $ & $ -0.170 \pm 0.069 $\\ 
 $ [15,17] $ & $ 0.00087\pm 0.0002$ & $0.071^{+0.001}_{-0.001}$
 			 & $ 0.12 \pm 0.15 $ & \\ 
 $ [17,18.9] $ & $ 0.00034\pm 0.00007$ & $0.027^{+0.001}_{-0.001}$
 			 & $ 0.20 \pm 0.26 $ &\\ 
 $ [1,6] $ & $ 0.0065\pm 0.0013$ & $0.985^{+0.020}_{-0.022}$
 			 & $ -0.03 \pm 0.14 $ & \\ 
 $[1.1,6]$ & $0.0064\pm 0.0013$   & 
&   &$0.020\pm 0.059$\\
 $[15,18.9]$ & $0.00066\pm 0.00013$ & $0.054^{+0.001}_{-0.001}$
            & $0.13\pm 0.11$ &$0.003\pm 0.046$\\ 
 %
\hline
$ A_5 $ & CCQM & PQCD~\cite{Jin:2020qfp} 
 & LHCb~\cite{Aaij:2015esa} & LHCb~\cite{LHCb:2021xxq}\\ 
$[0.1,0.98]$ & $0.268\pm 0.054$ & 
         && $ 0.043 \pm 0.067 $ \\
 $[0.1,2]$ & $0.0031\pm 0.0006$  &
             & $-0.02 \pm 0.13  $ & \\ 
$[1.1,4]$ & $0.0004\pm 0.0001$ &&  
& $ 0.026 \pm 0.067 $ \\

 $[2,5]$   & $-0.035\pm 0.007$   & 
         & $0.09\pm 0.25$  &   \\ 

$[4,6]$ & $-0.252\pm 0.050$ &
& & $-0.084 \pm 0.084 $ \\

 $[5,8]$ & $-0.3461\pm 0.0692$   &
         & $0.04\pm 0.17$     &   \\ 
 $[6,8]$ & $-0.3538\pm 0.0708$   & 
&     & $-0.022 \pm 0.082 $   \\

 $ [11,12.5] $ & $-0.33\pm 0.07$ & 
  				 & $ 0.08 \pm 0.21 $ &$ 0.035 \pm 0.063 $ \\ 
 $ [15,17] $ & $-0.25\pm 0.05$  & 
 				 & $ 0.02 \pm 0.14 $ &\\ 
 $ [17,18.9] $ & $-0.16\pm 0.03$  &
 				& $ 0.13 \pm 0.28 $& \\ 
 $ [1,6] $ & $ -0.034\pm 0.007$ & 
& $ 0.20 \pm 0.13 $& \\

 $[1.1,6]$ & $-0.108\pm 0.022$   & 
&  & $-0.007 \pm 0.051$ \\
 $[15,18.9]$ & $-0.21\pm 0.04$     &
         & $0.11\pm 0.10$     & $-0.025\pm 0.043$     \\ 
 %
\hline
$ A_8 $  & CCQM & PQCD~\cite{Jin:2020qfp} 
 & LHCb~\cite{Aaij:2015esa}& LHCb~\cite{LHCb:2021xxq}\\ 
$[0.1,0.98]$ & $0.0043\pm 0.0008$ & 
&& $ -0.007 \pm 0.073 $ \\

 $[0.1,2]$ & $-0.038\pm 0.008$   & 
            & $0.10 \pm 0.14 $ &\\ 
$[1.1,4]$ & $0.0043\pm 0.0008$ &&  
& $ 0.038 \pm 0.082 $ \\
 $[2,5]$   & $0.19\pm 0.04$      & 
               & $0.19\pm 0.24$&  \\ 
$[4,6]$ & $0.0017\pm 0.0003$ &
& & $0.012 \pm 0.090 $ \\

 $[5,8]$  & $0.0026\pm 0.0005$   & 
            & $-0.12\pm 0.18$ & \\ 
 $[6,8]$ & $0.0025\pm 0.0005$   & 
&     & $-0.170 \pm 0.080 $   \\

 $ [11,12.5] $ & $-0.0007\pm 0.0001$ &
 				 & $ -0.01 \pm 0.15 $ &$ 0.046 \pm 0.070 $ \\ 
 $ [15,17] $ & $-0.0003\pm 0.0001$ &
 				 & $ 0.08 \pm 0.17 $&\\ 
 $ [17,18.9] $ & $-0.0002\pm 0.0001$ &
 				 & $ -0.16 \pm 0.27 $&\\ 
 $ [1,6] $ & $0.17\pm 0.03$ & 
 				 & $ 0.00 \pm 0.16 $& \\ 

 $[1.1,6]$ & $0.0032\pm 0.0001$   & 
&  & $0.016 \pm 0.062$ \\

 $[15,18.9]$ & $-0.0003\pm 0.0001$      & 
              & $0.03\pm 0.12$ &$0.072\pm 0.051$ \\ 
 %
\hline
$ A_9 $ & CCQM & PQCD~\cite{Jin:2020qfp} 
 & LHCb~\cite{Aaij:2015esa} & LHCb~\cite{LHCb:2021xxq}\\ 
$[0.1,0.98]$ & $-0.0003\pm 0.0001$ & 
&& $-0.030 \pm 0.079 $ \\

 $[0.1,2]$ & $0.0065\pm 0.0013$  & 
           &  $0.03\pm 0.14$  & \\ 
$[1.1,4]$ & $-0.0007\pm 0.0001$ &&  
& $ 0.020 \pm 0.068 $ \\

 $[2,5]$   & $0.0065\pm 0.0013$  & 
           &  $-0.13\pm 0.27$ & \\ 
$[4,6]$ & $-0.0005\pm 0.0001$ &
& & $-0.008 \pm 0.061 $ \\

 $[5,8]$   & $-0.0010\pm 0.0002$  &
&  $-0.03\pm 0.17$ &\\
$[6,8]$   & $-0.010 \pm 0.0002$  & 
            &  & $-0.012\pm 0.090$\\ 
 $ [11,12.5] $ & $ 0.0005\pm 0.0001$ &
 			 & $-0.02 \pm 0.16 $ & $ 0.017 \pm 0.071 $\\ 
 $ [15,17] $ & $ 0.0004\pm 0.0001$ & 
 			 & $ 0.21 \pm 0.15 $ & \\ 
 $ [17,18.9] $ & $ 0.0002\pm 0.0001$ &
 			 & $ -0.02 \pm 0.19 $ &\\ 
 $ [1,6] $ & $ 0.0065\pm 0.0013$ & 
 			 & $ -0.01 \pm 0.13 $ & \\ 
 $[1.1,6]$ & $-0.0006\pm 0.0001$   & 
&   &$0.009\pm 0.046$\\
 $[15,18.9]$ & $0.0003\pm 0.0001$ &
            & $0.12\pm 0.10$ &$0.021\pm 0.042$\\ 
 %
\hline
\hline
\end{longtable}

\section{Discussion and Conclusion}

In this paper was shown the $q^2$-dependence of the $A_{i}$ and $S_{i}$ angular observables for $B_s\to\phi \ell^+\ell^-$ decay in Figs.~\ref{fig:P14}-\ref{fig:P15}, here $\ell$ is $\mu$ or $\tau$.
We have calculated the average values of the angular observables $<A_{FB}>$, $<F_L>$, $<P_1>$, $<P_4'>$, $<S_3>$, $<S_4>$ integrated over the full $q^2$ range for the $\mu$ and $\tau$ modes, finding them to be in agreement with the corresponding PQCD predictions~\cite{Jin:2020qfp}, except the $<F_L>$ for tau mode.

CCQM predictions for the branching fractions and angular observables for various bins in comparison with PQCD and LCHb data are given in Table~\ref{tab:bin1}. It should be noted that the CCQM predictions in Table~\ref{tab:bin1} are given for two-loop calculations for small and large regions of the $q^2$. Only values for bins $[5,8]$ and $[6,8]$ GeV$^2$ are given for one-loop calculations, due to the restrictions obtained in \cite{Greub:2008cy}.The inclusion of Next-to-next-to-leading logarithmic corrections to the Wilson coefficients $C_{7}^{eff}$  and $C_{9}^{eff}$ lead to non-zero values for observables such as $S_7$, which vanish at leading order. 

Generally CCQM predictions are consistent with data from LHCb collaboration~\cite{Aaij:2015esa,LHCb:2021zwz,LHCb:2021xxq} within the error bars. 
Predictions for branching ratios and longitudinal polarization fraction $F_L$ within CCQM are consistent with experimental data from 2021\cite{LHCb:2021xxq}, and the discrepancies do not exceed 1~$\sigma$.
The largest discrepancy is observed in the forward-backward asymmetry $A_{FB}$ for bins [6,8]~GeV$^2$ due to the reversed signs. The discrepancy for bins [11,12.5] and [15,18.9]~GeV$^2$ is about 4~$\sigma$.
For angular observable $A_5$, the largest discrepancies in bins [11,12.5], [15,18.9], and [6,8]~GeV$^2$ are about 3~$\sigma$.
The remaining angular observables do not exceed 2~$\sigma$ discrepancies.
It is noteworthy that discrepancies for most observables occurs in bins corresponding to [6-8]~GeV$^2$, where two-loop corrections cannot be used.

\section{ACKNOWLEDGEMENTS}
This work was supported by the Committee of Science of the Ministry of Science and Higher Education of the Republic of Kazakhstan (Grant No. AP19680084).

\end{document}